\documentclass[fleqn,10pt]{wlscirep}
\usepackage[utf8]{inputenc}
\usepackage[T1]{fontenc}
\usepackage{lineno}

\title{Global soil moisture from in-situ measurements using machine learning -- \textit{SoMo.ml}}

\author[1,*]{Sungmin O}
\author[1]{Rene Orth}
\affil[1]{Max Planck Institute for Biogeochemistry, Jena, D-07745, Germany}

\affil[*]{corresponding author(s): Sungmin O (sungmino@bgc-jena.mpg.de)}

\begin{abstract}
While soil moisture information is essential for a wide range of hydrologic and climate applications, spatially-continuous soil moisture data is only available from satellite observations or model simulations. Here we present a global, long-term dataset of soil moisture generated from in-situ measurements using machine learning, \textit{SoMo.ml}. We train a Long Short-Term Memory (LSTM) model to extrapolate daily soil moisture dynamics in space and in time, based on in-situ data collected from more than 1,000 stations across the globe. \textit{SoMo.ml} provides multi-layer soil moisture data (0–10 cm, 10–30 cm, and 30–50 cm) at 0.25$^{\circ}$ spatial and daily temporal resolution over the period 2000–2019. The performance of the resulting dataset is evaluated through cross validation and inter-comparison with existing soil moisture datasets. \textit{SoMo.ml} performs especially well in terms of temporal dynamics, making it particularly useful for applications requiring time-varying soil moisture, such as anomaly detection and memory analyses. SoMo.ml complements the existing suite of modelled and satellite-based datasets given its independent and novel derivation, to support large-scale hydrological, meteorological, and ecological analyses.
\end{abstract}
\begin{document}

\flushbottom
\maketitle

\thispagestyle{empty}

\section*{Background \& Summary}

\begin{figure}[ht]
\centering
\includegraphics[width=\linewidth]{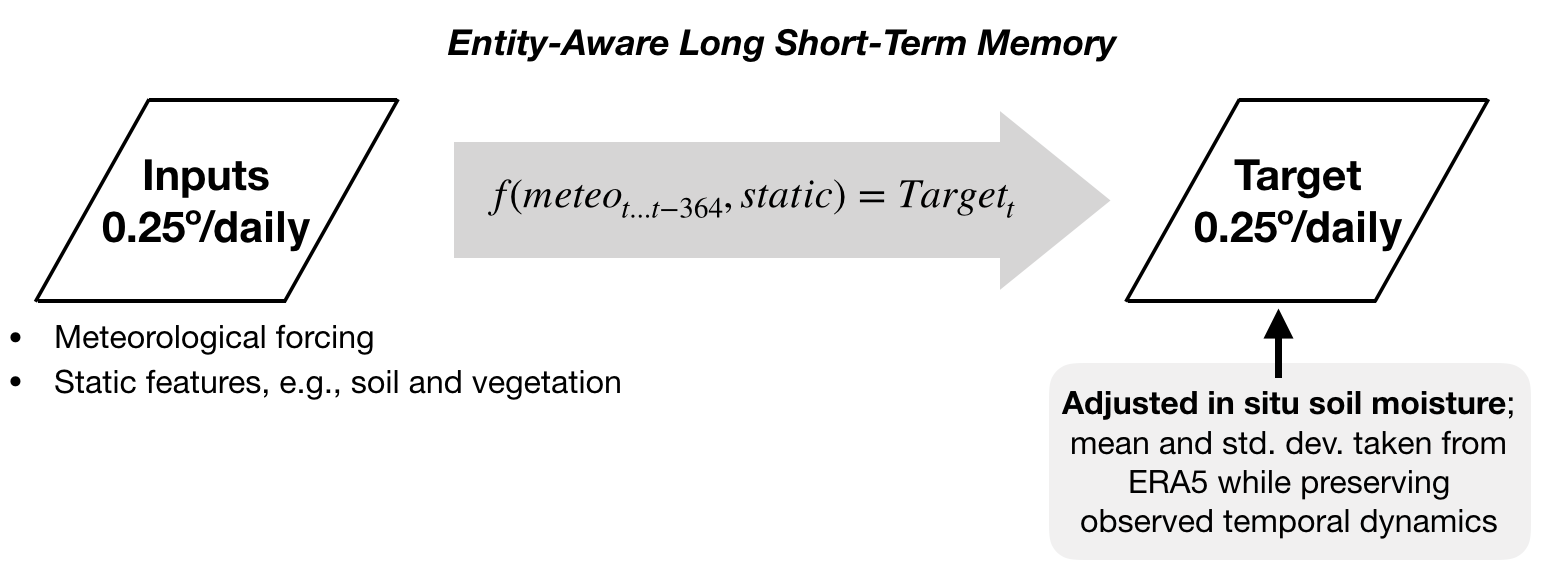}
\caption{Schematic of data-driven approach to generate global-scale gridded soil moisture from in-situ measurements}
\label{fig:fig1}
\end{figure}

Soil moisture plays a key role in land-atmosphere interactions through its control on water, energy, and carbon cycles \cite{daly_review_2005, seneviratne_investigating_2010}. Weather and climate variations are mediated by the soil moisture status \cite{koster_realistic_2004, orth_using_2014, prodhomme_impact_2016, jasper_2020}. Therefore, the spatiotemporal variations of soil moisture can influence the development and the persistence of extreme weather events such as heat waves, droughts, floods, and fires \cite{lorenz_persistence_2010, mueller_hot_2012, whan_impact_2015,sharma_if_2018, o_observational_2020}. For these reasons, soil moisture information is required to support a wide range of research and applications, e.g. agricultural monitoring, flood and drought prediction, climate projections, and carbon cycle modelling \cite{brown_nasas_2013}. Consequently, soil moisture is recognised as an Essential Climate Variable by the Global Climate Observing System \cite{gcos_2011}.\\ \\
Despite its scientific and societal importance, large-scale long-term observations of soil moisture are scarce. There is a significant number of in-situ soil moisture measurement networks \cite{dorigo_international_2011}, but they are not uniformly distributed. Satellite observations allow the derivation of global-scale soil moisture estimates; however, they represent only the top few centimetres of the soil. Moreover, satellite retrievals in areas with complex topography, dense vegetation, and frozen or snow-covered soils are challenging, leading to data gaps \cite{dorigo_esa_2017}. On the other hand, physically-based models can provide seamless soil moisture data at the global scale, but large differences exist across the models due to different and uncertain parameterisations of e.g. the spatial heterogeneity of soils and vegetation, and the non-linear relationship between soil moisture and evapotranspiration \cite{dirmeyer_gswp-2_2006, koster_nature_2009}. In summary, each source of soil moisture data has characteristic strengths and weaknesses. \\ \\
Meanwhile, machine-learning (ML) presents an alternative opportunity to produce seamless soil moisture data. The usefulness of ML algorithms for soil moisture estimation or forecasting has been demonstrated in the previous studies \cite{gill_soil_2006, ahmad_estimating_2010, adeyemi_dynamic_2018, fang_near-real-time_2019}. ML algorithms are able to ‘learn’ the complex relationship between soil moisture (target) and meteorological variables (predictors) from training data. In this way, soil moisture information can be inferred from readily observed predictor data in an empirical way without explicit knowledge of the physical behaviour of the system (e.g. land surface processes). The resulting soil moisture data is independent from, and can complement existing satellite-based or model-derived datasets. Similar data-driven approaches to derive gridded datasets using ML algorithms have been successfully employed in the cases of land-atmosphere fluxes \cite{jung_fluxcom_2019} and runoff \cite{ghiggi_grun_2019}.\\ \\
Here we present a novel global-scale gridded soil moisture dataset generated through a data-driven approach (Fig. \ref{fig:fig1}). Namely, we employ a Long Short-Term Memory neural network (LSTM) \cite{hochreiter_long_1997} to build a soil moisture simulation model. Daily meteorological time series and static features obtained from both reanalysis and remote sensing datasets are used as predictor variables. As a target variable, we use adjusted in-situ soil moisture measurements from different depths obtained from the International Soil Moisture Network (ISMN) \cite{dorigo_international_2011} and the National Center for Monitoring and Early Warning of Natural Disasters of Brazil (CEMADEN) \cite{zeri_soil_2020}. The raw point-level data are scaled to match respective means and variabilities obtained from European Centre for Medium-Range Weather Forecasts (ECMWF) ERA5 gridded soil moisture to allow seamless merging of measurements across different stations and time periods, and to estimate soil moisture at a target grid-scale.\\ \\
Our new global soil moisture dataset, \textit{SoMo.ml}, provides soil moisture at three different depths: 0-10 cm, 10-30 cm, and 30-50 cm, corresponding to Layer 1, Layer 2, and Layer 3, respectively. The data has a spatiotemporal resolution of 0.25$^{\circ}$ and daily, covering the period of 2000 to 2019. See Table \ref{Tab:table1} for more details.

\section*{Methods}
\subsection*{Target soil moisture data preparation}
\begin{figure}[ht]
\centering
\includegraphics[width=\linewidth]{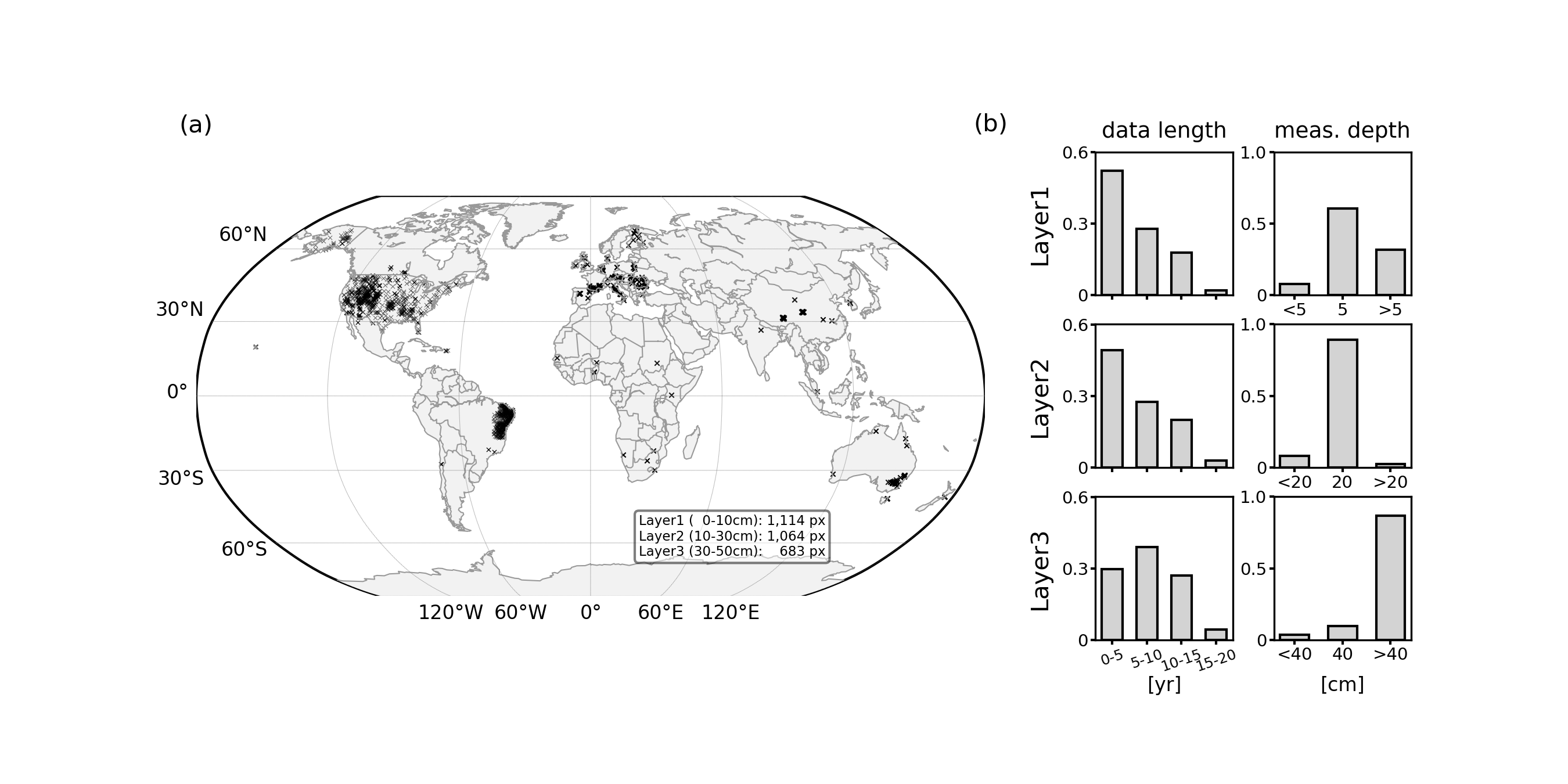}
\caption{(a) Spatial distribution of the target soil moisture data; 1114, 1064, and 683 grid cells are available for the layers of 0–10 cm, 10–30 cm, and 30–50 cm, respectively. (b) Data length and measurement depths of the target soil moisture over the period of 2000–2019.}
\label{fig:fig2}
\end{figure}

\begin{figure}[ht]
\centering
\includegraphics[width=\linewidth]{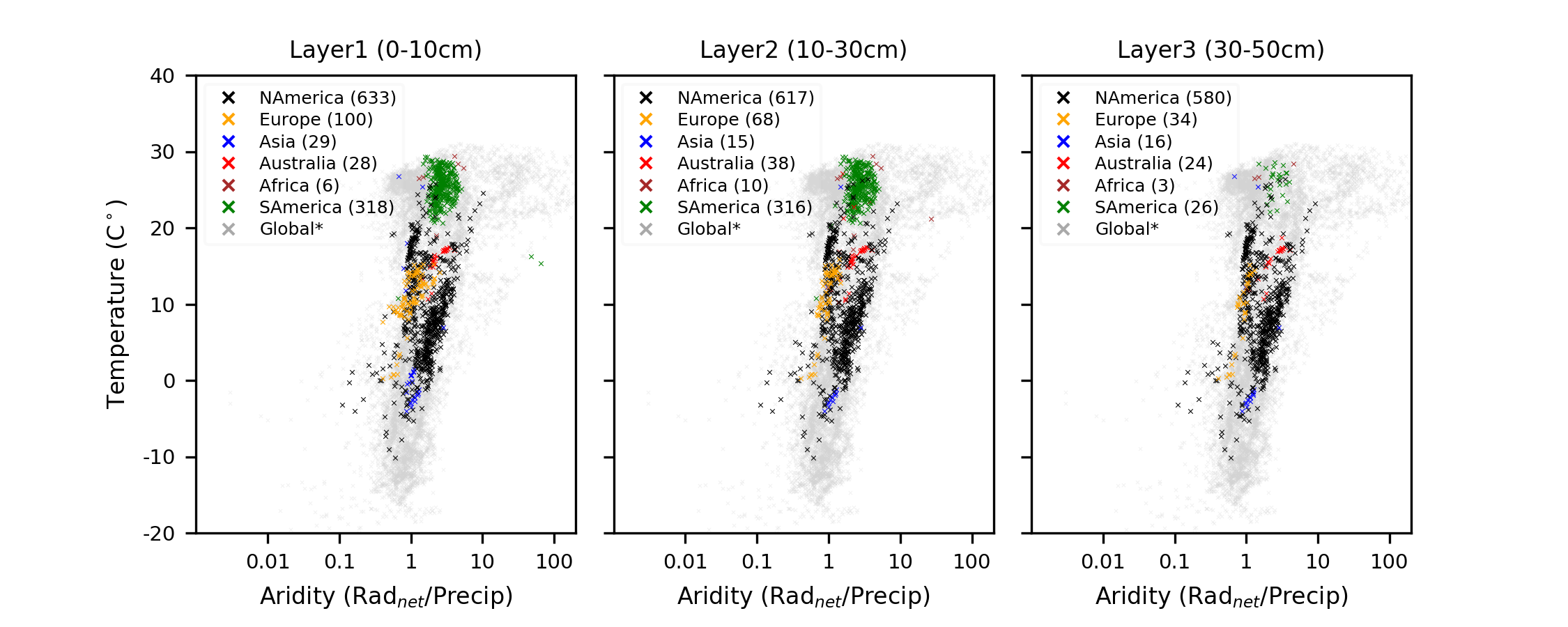}
\caption{Distribution of target soil moisture across hydro-climatic regimes for each layer. The total number of target data grid cells is given for each continent. Global grid pixels are randomly sampled (5 \%) from all land pixels for brevity.}
\label{fig:fig3}
\end{figure}

Target soil moisture data at 0.25$^{\circ}$ and daily resolution for model training is constructed using the in-situ measurements. From the ISMN data only ‘good’ observations are selected, based on the quality flag \cite{dorigo_quality_2013}. CEMADEN provides only useful-quality data \cite{zeri_tools_2018}. Both datasets provide sub-daily data and daily averages are computed for the days with at least six available sub-daily estimates. Stations or sensors with less than 2 months of data are discarded.\\\\
In-situ measurements across the different sites are collected with various sensor types, which have different calibrations. Therefore, the means and variances of the obtained time series are not necessarily comparable, which could introduce artifacts during the LSTM training. For this reason, we adjust the mean and standard deviation of the daily in-situ time series to those of the respective ERA5 grid-cell soil moisture within the overlapping period. As ERA5 soil moisture is available at 0-7 cm, 7-28 cm, and 28-100 cm depths, it is vertically interpolated into the target layer depths with a depth-weighted averaging. If more than one in-situ measurement time series is available at the same depth within the same grid cell (0.25$^{\circ}$), their average is taken. As a result, the adjusted in-situ target data resembles ERA5 soil moisture in terms of mean and standard deviation, while its daily temporal variations follow the ground observations. Our approach is also based on the fact that temporal variations from point-level data have a greater areal representation compared to absolute soil moisture values \cite{mittelbach_new_2012, malicke_soil_2020}. We can therefore assume that point-level data contains sufficient information to infer soil moisture dynamics at the grid scale.\\\\
For each soil layer, we preferentially select the adjusted in-situ measurement taken at the mid-depth of the layer; i.e. 5 cm, 20 cm, and 40 cm, respectively. If no data is available at the mid-depth, the measurement taken closest to the mid-depth, and within the layer, is chosen, leading to a total of 1114, 1064, and 683 grid pixels for the three layers, respectively. The location of the grid cells with available target soil moisture is shown in Fig. \ref{fig:fig2}a. Selected depths and data lengths of target soil moisture data employed for each layer are depicted in Fig. \ref{fig:fig2}b. A considerable fraction of the target data is obtained from North America across diverse hydro-climatic regions (see Fig. \ref{fig:fig3}). While training data from South America represents warm and semiarid regions, those from Asia mostly cover relatively cold regions. 

\subsection*{Model training}
LSTM is a special kind of recurrent neural networks that is capable of learning long-term dependencies across time steps in sequential data \cite{hochreiter_long_1997}. It has been widely used in land surface modelling such as runoff or soil moisture simulations \cite{adeyemi_dynamic_2018, fang_near-real-time_2019, kratzert_towards_2019, o_robustness_2020}. An adapted version of the LSTM architecture, \textit{Entity-Aware LSTM}\cite{kratzert_towards_2019}, that can ingest time-varying forcing and static inputs separately is used in this study, thereby allowing the algorithm to explicitly differentiate the two different types of information.\\\\
We model soil moisture using the Entity-Aware LSTM architecture (hereafter referred to as ‘LSTM model'); the model consists of 1) 128 of hidden units, 2) one LSTM layer with one dense layer, and 3) 0.5 of dropout rate. More details about the model validation can be found in Supplementary Information (see Figs. S1 and S2). The LSTM model is trained to learn the relationship between the multiple predictor variables and the target soil moisture. The model is trained separately for each soil layer. The predictor data used for the LSTM-based soil moisture modelling is listed in Table \ref{Tab:table2}. The meteorological inputs during days \textit{t-364} to \textit{t} are used to simulate soil moisture at day \textit{t}; i.e. the model can establish the relationship of present soil moisture with present and past meteorological forcing over a full annual cycle. All input data are normalised using their mean and standard deviation to enhance the training efficiency \cite{lecun_efficient_2012}. We use the mean squared error divided by the standard deviation of soil moisture at each individual grid cell as a loss function. This scaling ensures comparative values of the loss function across wet and dry regions with potentially different temporal variabilities \cite{kratzert_towards_2019}.\\\\
Commonly used meteorological forcing variables \cite{sheffield_development_2006, balsamo_revised_2009} are prepared from new global atmospheric reanalysis ERA5 produced by ECMWF \cite{hersbach_era5_2020}, as listed in Table \ref{Tab:table2}. For the deeper layers, soil moisture simulated from the upper layer(s) is further used as input data. ERA5 uses large amounts and diverse kinds of observations such as synoptic station data, satellite radiances, and ground-based radar precipitation information via the 4D-Var data assimilation. Its enhanced quality as meteorological forcing, compared to its predecessor ERA-Interim, has been demonstrated through an experiment with land surface models \cite{albergel_era-5_2018}. ERA5 is available in near real-time, allowing corresponding future updates of the \textit{SoMo.ml} dataset.\\\\
For the static data, long-term mean precipitation and aridity over the period of 2000-2019 is computed using the ERA5 data \cite{hersbach_era5_2020}. Aridity is defined as the ratio of net radiation (converted into ${mm}$) divided by precipitation \cite{budyko_1974}. We characterise topography through mean and standard deviation of sub-grid scale elevation, as obtained from the ETOPO1 digital elevation model \cite{amante_etopo1_2009}. In addition, we use soil type and land cover information from the Global Land Data Assimilation System (GLDAS) data archive \cite{rodell_global_2004}. GLDAS resampled soil porosity and fractions of sand, silt, and clay from FAO datasets \cite{reynolds_estimating_2000} into 0.25$^{\circ}$ spatial resolution. The land cover is based on MODIS-derived 20-category vegetation data that uses a modified International Geosphere–Biosphere Programme classification scheme \cite{friedl_global_2002}. We use GLDAS Dominant Vegetation Type Data Version 2 which assigned the predominant vegetation type to each 0.25$^{\circ}$ grid cell.
 
\subsection*{Global data generation} 
The LSTM model is trained using the entire training dataset which consists of the available target soil moisture data and corresponding predictor data. After establishing the internal relationships (‘learning’), the model is applied using the predictor data over a quasi-global area of 90$^{\circ}$–60$^{\circ}$ at 0.25$^{\circ}$ spatial resolution. In order to account for the random initialisation of LSTM's trainable parameters, five simulations are performed and final soil moisture values are computed as an average of the five simulations.

\section*{Data Records}
The \textit{SoMo.ml} dataset can be accessed at \url{https://figshare.com}. Three compressed files (.zip) contain data in NetCDF format for the three respective layers. An example file name is 'SoMo.ml\_v1\_<LAYER>\_<YYYY>.nc', with LAYER and YYYY standing for soil moisture layer depth and year, respectively.

\section*{Technical Validation}
\subsection*{Comparison with independent in-situ measurements}
\begin{figure}[ht]
\centering
\includegraphics[width=12cm]{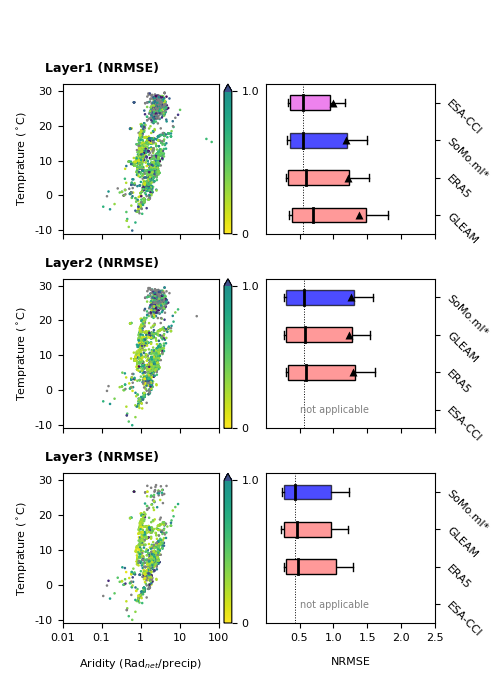}
\caption{Comparison of absolute soil moisture between \textit{SoMo.ml*} and in-situ data for each layer (top to bottom): (left) NRMSE values of \textit{SoMo.ml*} at each measurement station and (right) comparison with other global gridded datasets. Box plot whiskers show the 0.2 to 0.8 quantiles of the NRMSE across all measurement stations. The boxes are ranked according to the median NRMSE so that the best performing data is positioned at the top.}
\label{fig:fig4}
\end{figure}

\begin{figure}[ht]
\centering
\includegraphics[width=12cm]{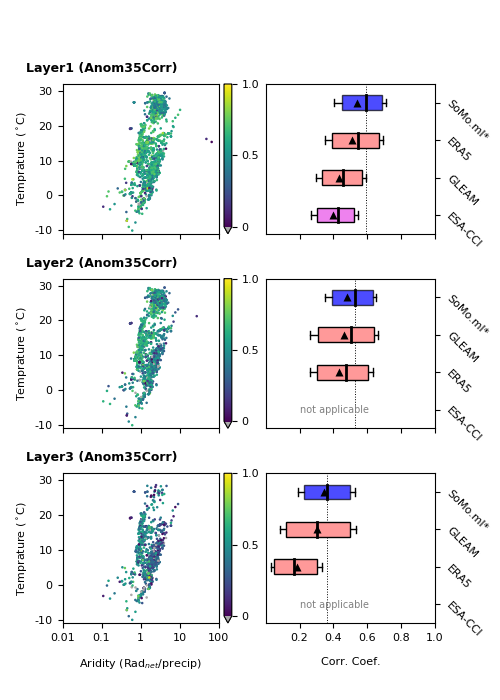}
\caption{Same as in Fig. \ref{fig:fig4}, but for correlation coefficient of anomalies where anomalies are determined by removing the mean of a surrounding 35-day window for each value.}
\label{fig:fig5}
\end{figure}

Cross-validation (5-fold) is made through a direct grid-to-point comparison between the \textit{SoMo.ml*} and the in-situ measurements as done in many previous studies \cite{albergel_soil_2012, martens_gleam_2017, pablos_assessment_2018, al-yaari_validation_2018, li_comprehensive_2020}. The simulated soil moisture for the validation is hereafter referred to as \textit{SoMo.ml*}, as this simulation data differs somewhat from the actual \textit{SoMo.ml} because it is not based on training with all available target data, but only with 80\% of the data according to the 5-fold cross validation approach. This validation also enables a comparative assessment of modelled soil moisture from the LSTM with that of state-of-the-art global gridded datasets such as ERA5, GLEAM\cite{martens_gleam_2017}, and the satellite-based ESA-CCI \cite{dorigo_esa_2017} datasets. Established skill scores such as normalised root-mean-square error (NRMSE), relative bias, and correlation coefficient are used to quantify the agreement with the ground truth data.\\\\
Figure \ref{fig:fig4} shows the distribution of the NRMSE of \textit{SoMo.ml*} across climate regimes (left) and a comparison of these results with the respective performances of the reference datasets (right). NRMSE is defined as the RMSE divided by the means of ground truth. Although \textit{SoMo.ml*} shows slightly higher biases at some stations over warm and arid regions, there is no clear overall climate dependency of the NRMSE. In Layer 1, while the median NRMSE of \textit{SoMo.ml*} is similar to that of ESA-CCI, which shows lowest NRMSE, a wider spread of errors is observed. ERA5 and GLEAM tend to overestimate in-situ measurements (see Fig. S3 in Supplementary Information for relative biases), leading to slightly higher NRMSE values. In the deeper layers, where ESA-CCI is not available, NRMSE values of \textit{SoMo.ml*} are slightly lower but overall similar to those of the ERA5 and GLEAM references. As a result, this comparison highlights similar deviations of absolute soil moisture values from in-situ measurements across the considered datasets.\\\\
Figure \ref{fig:fig5} shows results from a similar comparison, but focusing on the time-variability of the soil moisture dataset as expressed by the correlation of soil moisture anomalies with in-situ measurements. To exclude the impact of the seasonal cycle, we consider short-term anomalies\cite{albergel_skill_2013, dorigo_evaluation_2015}. For each soil moisture at day \textit{d}, a period \textit{P} is defined as \textit{P}= [d-17, d+17] (corresponding to a 5-week window). If at least 10 data are available within the period, the average soil moisture and corresponding anomaly are computed. Equations are applied to each station and a grid pixel it lies on. No pronounced climate dependency of the correlations is observed for \textit{SoMo.ml*} (Fig. \ref{fig:fig5}, left). Comparing with the reference datasets, \textit{SoMo.ml*} outperforms them for the top layer. While overall anomaly correlations decrease in the deeper layers, also for these layers \textit{SoMo.ml*} shows closer agreement with the observations than the reference datasets. The results underline the particular strength of \textit{SoMo.ml*}, and likely also the actual \textit{SoMo.ml}, to represent the temporal variability of soil moisture. This is somewhat expected; while this comparison is done against independent in-situ measurements, the temporal dynamics of \textit{SoMo.ml*} are directly learned from (remaining) in-situ measurements. Similar results are obtained when using the correlations of long-term absolute soil moisture, and of anomalies derived by removing the mean daily averages (Figs. S4 and S5, respectively).\\\\
Note that ESA-CCI has missing values in space and time and GLEAM is available only until 2018, such that partly different spatiotemporal data are used among datasets in the comparison. We repeat the analysis above using only data where all datasets are available and find very similar results (not shown). In summary, compared with state-of-the-art references, \textit{SoMo.ml*} shows a comparable performance in terms of biases, while outperforming the other datasets in terms of temporal correlations.

\subsection*{Global-scale comparison with existing gridded datasets}
\begin{figure}[ht]
\centering
\includegraphics[width=\linewidth]{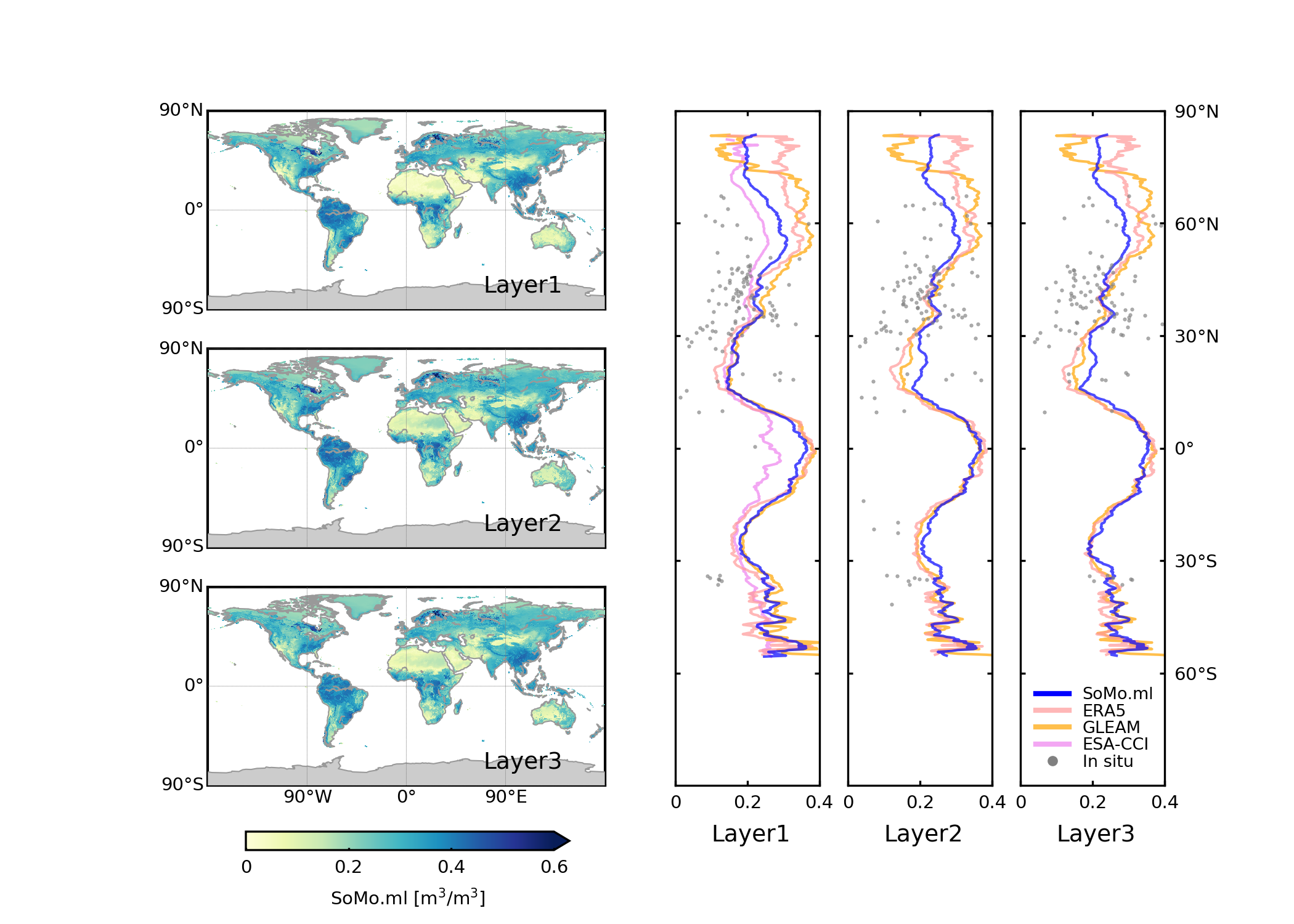}
\caption{(a) Global maps of 20-year long-term medians of \textit{SoMo.ml}. (b) Comparison of latitudinal median profiles among the considered datasets. In the case of GLEAM, root-zone soil moisture is used for both Layer 2 and Layer 3.}
\label{fig:fig6}
\end{figure}

Next, we examine the spatial patterns of \textit{SoMo.ml} at the global scale. Figure \ref{fig:fig6}a presents the median soil moisture values over the entire period. Low values in arid regions such as southwest North America, North Africa, central Asia, and Australia and high values in more humid regions such as the northern latitudes and Southeast Asia are well captured. A comparison of the distributions of global median soil moisture between \textit{SoMo.ml} and the references can be found from Figs. S6 to S8. Figure \ref{fig:fig6}b compares latitudinal profile of \textit{SoMo.ml} against that of the reference datasets (Fig.\ref{fig:fig6}b). Overall, we find a satisfactory consistency between global patterns of \textit{SoMo.ml} and the reference datasets. For instance, the highest average soil moisture occurs near the equator in the tropics, while driest soil moisture is found near 20$^{\circ}${N}. These patterns are overall well reproduced in \textit{SoMo.ml}. This is expected to some extent because we rescale the target soil moisture using ERA5 means and standard deviations, such that the LSTM algorithm will pick up these ERA5 characteristics in locations and at time steps with available in-situ measurements. Nonetheless, \textit{SoMo.ml} between 15$^{\circ}${N} and 25$^{\circ}${N} tends to be wetter than the reference datasets (over the eastern part of the Sahara desert), especially in the deeper layers. More generally, \textit{SoMo.ml} might not properly describe soil moisture in very-arid regions, which can be related to a lack of training data from such regions (see Fig. \ref{fig:fig3}). Different patterns found in ESA-CCI are likely due to the missing data. Over very high latitudes over 60$^{\circ}${N}, we can observe relatively large differences across datasets, probably due to different freezing and thawing patterns. Meanwhile, in-situ measurements (not adjusted) do not show a meaningful pattern of latitudinal averages but large variability across stations and sensors, whereby it is not clear to which extent this is due to different sensor types and calibrations or due to actual moisture differences caused by heterogeneous land surface characteristics.

\section*{Usage Notes}
We present a global, multi-layer, long-term soil moisture dataset generated through a novel data-driven approach, and with comprehensive ground truth data. For model training, we preprocess the in-situ measurements to obtain more spatiotemporally consistent, grid-scale target soil moisture data by adopting mean and standard deviation from ERA5 data while preserving the observed temporal variations from the in-situ measurements. At the same time, our newly generated soil moisture data outperforms other existing gridded datasets, including ERA5, in terms of daily temporal dynamics as indicated by highest temporal(anomaly) correlation with the ground observations. Nonetheless, the data quality in conditions outside the spatiotemporal range sampled within the observations is potentially uncertain. LSTM performance can be significantly affected by the (lack of) hydro-climatic diversity in the training data, even more than by the quantity of data \cite{o_robustness_2020}. As shown in Fig. \ref{fig:fig3}, while the in-situ soil moisture measurements are obtained from networks worldwide, the data does not cover all globally occurring hydro-climatic conditions. Therefore, relatively high uncertainty outside the training conditions such as at high latitudes and in arid regions is expected. However, this lack of observations in particular conditions also presents a challenge to other datasets/models \cite{dorigo_evaluation_2015, reichle_smap_2017}. Therefore, for instance, using \textit{SoMo.ml} within an ensemble of independently derived datasets could be a promising solution to obtain more reliable soil moisture information in these data-sparse regions \cite{guo_improving_2007, wang_multimodel_2009}. As a result, our new independent soil moisture dataset is a valuable addition to the existing suite of soil moisture datasets, and can enhance future large-scale hydrologic and ecologic analyses, and also benchmark studies to evaluate land surface models and remote sensing data.\\\\

\section*{Code availability}
We implement the LSTM model by adopting python modules obtained from \url{https://github.com/kratzert/ealstm\_regional\_modeling}. Python codes to generate and/or analyse data are available from the corresponding author on reasonable request. 

\bibliography{references}

\section*{Acknowledgements}
We would like to thank Ulrich Weber (Max Planck Institute for Biogeochemistry) for preprocessing and providing the datasets. We also thank Dr. Sophia Walther (Max Planck Institute for Biogeochemistry) for her valuable comments. This study is supported by the German Research Foundation (Emmy Noether grant 391059971).

\section*{Author contributions statement}
SO and RO designed the study. SO performed the computations and data analysis. All authors discussed the results and wrote the paper.

\section*{Competing interests}
The authors declare that they have no conflict of interests.

\section*{Figures \& Tables}

\clearpage

\begin{table}[ht]
\centering
\begin{tabular}{|p{3.5cm}|p{12cm}|}
\hline
Data type & Gridded \\\hline
Spatial Extent & Quasi-global (90$^{\circ}$N–60$^{\circ}$S)\\
\hline
Temporal coverage & 2000 to 2019\\
\hline
Spatial Resolution & 0.25$^{\circ}$ x 0.25$^{\circ}$\\
\hline
Temporal Resolution & daily\\
\hline
Variables& Soil moisture at three layers (0–10 cm, 10–30 cm, and 30–50 cm)\\
\hline
Unit& Volumetric soil moisture [${m{^3} m^{-3}}$]\\
\hline
File format & NetCDF \\
\hline
Key strengths & 1) Global scale, long-term data.

2) Largely independent from available gridded soil moisture products.

3) Better agreement with in-situ measurements in terms of temporal soil moisture dynamics.\\\hline
Limitations&
1) Performance depends on in-situ data availability, which is low in tropical regions including Africa.

2) Uncertainty and errors in measurements may affect the model performance.

3) ERA5-based scaling is necessary, making means and variabilities of \textit{SoMo.ml} similar to ERA5 data.\\
\hline
\end{tabular}
\caption{\label{Tab:table1}Specifications of \textit{SoMo.ml} v1.}
\end{table}

\vspace*{100px}

\begin{table}[ht]
\centering
\begin{tabular}{|p{1.5cm}|p{5.25cm}|p{2cm}|p{5.75cm}| }
\hline
& Variable & Source & Description\\
\hline
Dynamic & Air temperature & ERA5\cite{hersbach_era5_2020} & Daily meteorological forcing obtained from ECMWF reanalysis\\
& Precipitation &  & \\
& Specific humidity &  & \\
& Net surface radiation & & \\
& Downward surface solar radiation & & \\
& Land surface temperature &  & \\
& Soil moisture from upper layer(s) for second and third layers & SoMo.ml\cite{somo_dat} & ML-based soil moisture produced in this study\\
\hline
Static & Mean precipitation & ERA5\cite{hersbach_era5_2020} & Long-term mean precipitation\\
& Aridity & ERA5\cite{hersbach_era5_2020} & Ratio of net radiation to precipitation\\
& Topography & ETOPO1\cite{amante_etopo1_2009} & Mean and standard deviation of sub-grid scale elevation values at each grid cell\\
& Vegetation type & GLDAS\cite{rodell_global_2004} & Predominant vegetation type (MODIS-derived) at each grid cell\\
& Soil type & GLDAS\cite{rodell_global_2004} & Clay, sand and silt fractions based on FAO Soil Map of the World\cite{reynolds_estimating_2000}\\
& Soil porosity & GLDAS\cite{rodell_global_2004} & Soil porosity across layers, based on FAO Soil Map of the World\cite{friedl_global_2002}\\
\hline
\end{tabular}
\caption{\label{Tab:table2}Predictor data used for the LSTM model.}
\end{table}

\end{document}